# Determination of Multipath Security Using Efficient Pattern Matching

James Obert*
Cyber R&D Solutions
Sandia National Labs
Albuquerque, NM, USA
jobert@sandia.gov

Huiping Cao
Computer Science Department
New Mexico State University
Las Cruces, NM, USA
hcao@cs.nmsu.edu

*Abstract*—Multipath routing is the use of multiple potential paths through a network in order to enhance fault tolerance, optimize bandwidth use, and improve security. Selecting data flow paths based on cost addresses performance issues but ignores security threats. Attackers can disrupt the data flows by attacking the links along the paths. Denial-of-service, remote exploitation, and other such attacks launched on any single link can severely limit throughput. Networks can be secured using a secure quality of service approach in which a sender disperses data along multiple secure paths. In this secure multi-path approach, a portion of the data from the sender is transmitted over each path and the receiver assembles the data fragments that arrive. One of the largest challenges in secure multipath routing is determining the security threat level along each path and providing a commensurate level of encryption along that path. The research presented explores the effects of real-world attack scenarios in systems, and gauges the threat levels along each path. Optimal sampling and compression of network data is provided via compressed sensing. The probability of the presence of specific attack signatures along a network path is determined using machine learning techniques. Using these probabilities, information assurance levels are derived such that security measures along vulnerable paths are increased.

## I. INTRODUCTION

Typical network protocols select the least-cost path for routing data to destinations and thus address delivery efficiency along a single network path. On networks using single-path routing, attackers can launch attacks upon any link which seriously compromises data integrity, availability, and confidentiality along the path. Network countermeasures required along a compromised path include TCP resets of the offending attack node or nodes and involves disrupting the flow of traffic on the path for a period of time, and switching to an alternate path. Nevertheless, deploying these countermeasures generally requires manual intervention and an associated switching time [1]. Having multiple paths available for traffic propagation hinders an attacker's ability to focus the attack on a single routing path. However, multipath traffic propagation conversely introduces complexity into the system: using multiple paths requires sophisticated packet-reordering methods and buffering methods [2], [3]. In a fully secure multipath network a sender simultaneously transmits data over multiple paths with varying levels of security enabled along each path. The level of security along each path should reflect a measured threat level on the path and be dynamically adjusted as the attack environment varies.

Despite the importance of associating and adjusting the security level to each path in multipath routing, existing multipath routing protocols such as Multipath TCP lack the ability to actively determine the level of security threats along a path [4], [31].

In this paper, we present a novel approach that utilizes compressed sensing (CS) [13] and machine learning techniques to determine the information assurance level of network paths in multipath networks. Compressed sensing (CS) allows network data to be optimally sampled below the normally required Nyquist 2X signal sampling frequency while simultaneously compressing data and lowering data dimensionality. CS data compression enables the storage of large data windows by up to a factor of 10X. The combination of data compression and data dimensionality reduction effectively filters out non-contributing network traffic features which increases the efficiency and data handling capabilities of anomaly detection algorithms used in network path security determination.

Compared to other types of multipath network security methods, the proposed approach is based on recognizing real-world attack patterns within compressed and dimension reduced data sets. Additionally, most multipath security schemes are based on hypothetically derived trust models while the proposed approach finds the likelihood of the presence of real-world attack patterns in data event windows and assigns information assurance levels to paths that can be subsequently utilized to actively adjust path security measures [1-8].

The remainder of this paper is organized as follows. We provide in Section II a review of related work. Section III presents the *compressed sensing - signature cluster path security determination* methods. In section IV evaluation results are presented, and finally in section V conclusions are discussed.

## II. BACKGROUND

In multipath routing, data is transmitted along multiple paths to prevent fixed unauthorized nodes from intercepting or injecting malicious data onto a network. Ideally, the simplest form of multipath routing entails using no encryption and data is split among different routes in order to minimize the effects of malicious nodes. The approach in [5] uses existing multiple paths such that an intruder needs to spread resources across several paths to seriously degrade data confidentiality. In the approach of [6], one path is used as a central path while the other paths are alternatives. When the central path's performance is seriously affected, one of the alternative paths is selected as the new central path. These two multipath protocols base the effectiveness on the ability to either disperse data along multiple paths or in having the option to switch to alternate paths. However, none of the approaches suggests an adequate or explicit means for combining dispersive data security methods with path differentiating data security measures.

The differentiating approach proposed in this paper is to intelligently sense the threat level present along each network path and correspondingly increase the encryption strength on more vulnerable paths while decreasing it on the less vulnerable ones. In order to maintain overall throughput, the transmission rates on more vulnerable paths will drop, while it will increase on the less vulnerable ones. The proportional multipath encryption and routing approach is expressed in Eq. (1) and maintains a *secure quality of service (SQoS)*. Packets are proportionally routed over paths $P_i$ and $P_j$ according to values $I, C, E$ over graph edges, which are defined shortly.

### III. NETWORK PATH SECURITY DETERMINATION

Given a network, let $I$ be the information assurance factor, $C$ be the link cost factor (i.e., OSPF cost), and $E$ be the encryption scaling factor. For distinct edges or links in a network, the values of these factors are different. To differentiate the factor values on different links, we use subscript $i$ to denote the factor value for an edge $e_i$. E.g., $I_i$ is the information assurance factor for an edge $e_i$. Given a message with length $L$, we need to formulate the throughput for sending this message from a source node $v_s$ to a destination node $v_e$ when using multipath routing by leveraging these factors. In general, if all paths that are used to send a message is $|P|$, and the length of a path $P_i$ is $n_i$, then the throughput is defined as follows.

$$T_{v_s \to v_e} = L \sum_{i=1}^{|P|} \sum_{j=1}^{n_i} I_{ij} C_{ij} E_{ij} \quad (1)$$

For example, assume that the network routing algorithm decides to use two paths $P_i$ = path ($v_1, v_6, v_3, v_4, v_2$) and $P_j$ = path ($v_1, v_6, v_5, v_7, v_2$) to send a message with length $L$ from $v_1$ to $v_2$ in Figure 1.

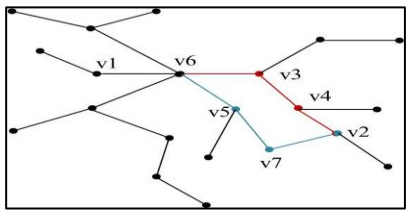

**Figure 1:** Multipath Graph

Then, its throughput is:

$$T_{v1 \to v2} = L \cdot \sum_{i=1}^{n_i} I_i C_i E_i + L \cdot \sum_{j=1}^{n_j} I_j C_j E_j \quad (2)$$

The throughput to destination vertex "$v_e$" is maintained, but the encryption "$E$" scaling factors are dynamically adjusted according to the values of the information assurance factor $I$ over each edge.

It will be shown that the information assurance factors $I$ along a path can be derived by finding the likelihood of the presence of attack signature patterns within a defined event window of network traffic (Section III.D). Link encryption factors $E$ and link cost factors $C$ are inversely proportional to the value of information assurance factors $I$. Derivation of factors $E$ and $C$ in maintaining SQoS and throughput $T_{v_s \to v_e}$ is reserved for future research.

Our approach determines the security levels of network paths by examining the traffic data with different temporal partitions. In particular, the network traffic is partitioned into event windows where each window collects data over 30 minute sampling periods. For each 30-minute event window, we collect $N$ sample from the network traffic for a single path.

For each event window, our approach performs traffic sampling, anomaly detection, and path security determination as shown in the diagram of Figure 2.

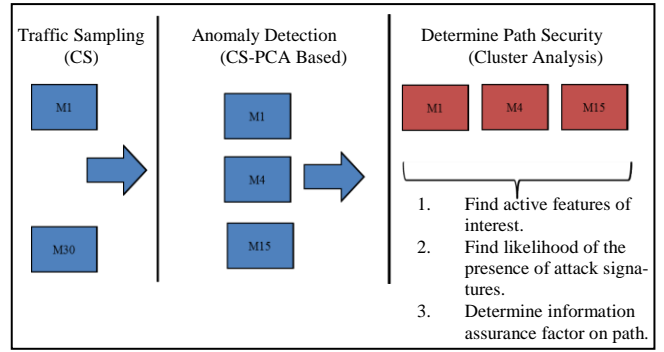

**Figure 2:** Processing Flow for Traffic in One Event Window

As Figure 2 shows, compressed sensing (CS) [13] is used to optimally sample network traffic data and store them in a compressed form (Section III.A). Behavioral anomaly detection is conducted on the CS data (Section III.B and Section III.C). The compressed data are passed to the path security determination process (Section III.D), which performs cluster analysis on the traffic samples. Significant clusters are inspected for the presence of active attack signature features, and the likelihood of a respective cluster containing attack signatures is calculated (Section III.D). Given the likelihood of specific attack features being present on a path $P_i$, the cyber threat level $W_r$ and information assurance $I_i$, are determined using Eq. (13) and Eq. (14), which are discussed in Section III.E. In what follows, we discuss every step of Figure 2 in detail.

### A. Traffic Sampling

Network packet header, network time, port, protocol, and flags are collected at each router interface. An event window corresponds to a set of TCP/IP packet records for a path transiting a set of subnets or virtual LANs (VLAN) contained within an autonomous system.

For each event window, we first define several notations used in the process of data sampling and the later discussions.
- $f$: one feature that is abstracted from the network packet records. When there are multiple features, we also use $f_i$ to denote the $i$-th feature.
- $N^f$: the total number of features that are of our interest. In this research, a total of 19 features were extracted from the captured network packets. These features correspond to a specific subset of TCP, ICMP, HTTP, OSPF protocol states which are most often associated with router and host attacks.
- $X_f$: samples for a feature $f$ in one event window. It records the number of samples of feature $f$ at different sample moment. If there are $N$ sample moments, then $X$ is an $N$–dimensional column vector.
- $N$: the number of samples that we take for one event window.
- $\Phi = (X_1, X_2, ... X_{Nf})$: a $N \times N^f$ matrix with the $N$ samples that are taken for an event window. Here, $X_i$ is represented as a $N$-dimensional column vector.

Three separate network based suits, namely reconnaissance, vulnerability scanning, and exploitation, were used in emulating real-world host and network conditions. Each suite possesses a unique *signature* ($S_r$). The *threat level* ($W_r$) is assigned to each attack suite type, which ranges from 1 for least severe to 5 for most severe. Table 1 shows the detailed information of the network attack suites that we used in this research.

**Table 1** Network Attack Suites

| Suite Signature ($S_r$) | Description | Active Features | Threat Level ($W_r$) |
|---|---|---|---|
| 1 | Cloud Guest Reconnaissance, Vulnerabilities & Exploitation | 6 $\{f_1, f_2, f_3, f_4, f_9, f_{11}\}$ | 3 |
| 2 | Cloud Infrastructure Reconnaissance, Vulnerabilities & Exploitation | 6 $\{f_1, f_5, f_6, f_7, f_8, f_9\}$ | 5 |
| 3 | Cloud Services Reconnaissance, Vulnerabilities & Exploitation | 5 $\{f_1, f_5, f_8, f_9, f_{10}\}$ | 4 |

Both compressed and uncompressed data in event windows were used in the analysis. Compressed data in event windows were created by sampling the uncompressed data in the corresponding event window.

### B. Data Compression Using Compressed Sensing

Compressed data for each event window are calculated using the CS technique [12, 13]. The theory of the CS technique is explained as follows.

Compressed sensing relations are listed below. For the observed data $x \in R^N$ with $Q$ representing the number of non-zero elements. The value of $Q$ is determined by finding those vectors where the sum of the absolute values of the columns is minimum. This otherwise known as the *L-1* norm and represented by $min_x \|x\|_1$.

$$min_x \|x\|_1 \text{ subject to } Y = U_v x \quad (3)$$

In Eq. (3) $U_v \in R^{M \times N}$ is an $M \times N$ orthogonal matrix called the sensing or measurement matrix, $v$ is a random subset of the row indices, and $Y$ is the linearly transformed compressed data.

We note that the $|V| = M$ and dictates the level of compression which is afforded when the linear transformation is performed on $\Phi$.

$$|V| \geq \text{Const} \cdot \mu^2(U) \cdot Q \cdot \log N \quad (4)$$

$$\mu(U) = max_{ij} |U_{ij}| \quad (5)$$

$Y = U_v x$ is a linear transformation reducing the data dimensionality from $N$ to $M$ with $U_v$ columns normalized to unit norm. If the sparseness of $x$ is considered, a dimension $k$ represents the components with high variance, and $M$ is chosen such that $M \geq k$.

From Eq. (4), the CS sampling rate which yields the best results is captured in Eq. (6) where $\varepsilon$ is a constant proportional to the number of active features and $M$ is the number of samples to be taken.

$$M = \varepsilon * \sqrt{N} * \log N \quad (6)$$

### C. Anomaly Detection

The previous step calculates the compressed data $Y$ from the original traffic data $\Phi$ for each event window. Our anomaly detection component detects the event windows that may contain traffic with anomalous behavior. In this section, we first describe the detailed steps of this component. Then, we explain the theory behind each step.

The anomaly detection component works as follows. The first step performs Principal Component Analysis (PCA) on the compressed data from one event window. I.e., PCA is performed on a covariance matrix of $Y$. The second step applies a residual analysis over the original data $\Phi$ and calculates a squared prediction error. If the prediction error for one feature is bigger than a threshold, then that event window is considered to contain anomalous behavior.

The compressed event window is represented by $Y$ in Eq. (7).

$$Y = U_v x \qquad (7)$$

The sampling matrix $U_v$ projects $x$ to a residual subspace; however, eigenvalue decomposition of the covariance matrix of $x$ and $Y$ yields near-identical eigenvalue magnitudes from which anomaly detection can be derived [10]. This fact allows one to inspect the compressed data samples, $Y$, for anomalies reducing the computational complexity to $\mathcal{O}(M^3)$ and storage of the data to $\mathcal{O}(M^2)$ with $M = \mathcal{O}(k \log N)$. This is a substantial running time reduction of the PCA analysis over the covariance matrix of $x$, which requires $\mathcal{O}(N^3)$ computations and memory storage of $\mathcal{O}(N^2)$.

A residual analysis method [9] decomposes the observed data $x$ (in our case, one row vector in $\Phi$) into principal subspace which is believed to govern the normal characteristics. Within the residual subspace in which $Y$ resides, abnormal characteristics can be found. The residual method performs the eigenvalue decomposition of the covariance matrix of $x$ from which $k$ principle eigenvectors $E$ are obtained. The projection of a data instance $x$ onto the residual subspace is

$$z = (I - EE^T)x \qquad (8)$$

Assuming the data is normal, the squared prediction error is $\|z\|_2^2$ which follows a non-central chi-square distribution. Anomalous activity is detected when the squared prediction error $\|z\|_2^2$ exceeds a certain threshold called $Q$-statistics which is the function of the non-principle eigenvalues in the residual subspace and is approximated by

$$Q_\beta = \theta_1 \left[ \frac{c_\beta \sqrt{2\theta_2 h_0^2}}{\theta_1} + \frac{\theta_2 h_0 (h_0-1)}{\theta_1^2} \right]^{\frac{1}{h_0}} \qquad (9)$$

where $h_0 = 1 - \frac{2\theta_1 \theta_3}{\theta_2^2}$, $\theta_i = \sum_{j=p}^N \lambda_j^i$ for $i = 1, 2, 3$, $c_\beta = (1-\beta)$ percentile in a standard normal distribution and $Q_\beta$, and $\lambda_j$, $i = 1,...,k$ are the eigenvalues of the covariance matrix. Anomalies are detected when the prediction error $\|z\|_2^2 > Q_\beta$. [9]

### D. Determination of Path Security

Using the approach discussed in Section III.C, volume-based anomalous behavior within an event window is identified. Such anomalous behavior provides an indication that an attack may exist within this event window. If anomalous behavior exists in an event window with high probability, then this component attempts to determine the security level for paths in this event window by using hierarchical clustering techniques described in this section.

Agglomerative hierarchical clustering was chosen as the method for deriving anomalous and baseline models because of its ability to identify clusters without providing an initial estimate of the number of clusters present. Agglomerative hierarchical clustering algorithm iteratively groups data points or clusters of points to form new clusters. Each iteration results in the previously found points and clusters being clustered with another point or cluster. Generally, the results of hierarchical clustering of sizable data sets are a large number of clusters, many of which contain a small fraction of the samples. A straightforward approach to prioritizing clusters is to eliminate the minor clusters by cutting the hierarchical dendrogram lower tiers.

Once clusters are identified in an event window (Algorithm 1), determination of which clusters contain attack signature features of high magnitude is conducted (Algorithm 1). The path information assurance factor is calculated (Algorithm 2).

In order to lower computational complexity, only those event windows found to have volumetric anomalies in the residual subspace are used in determining network path security. The relative magnitudes and spectral properties of each feature in principal subspace are calculated, and the uncompressed form of each anomalous event window is analyzed. A signature consisting of a distinct collection of significant features is associated with each attack suite; thus, the nature of significant features contained within the traffic data of an event window is captured in a hierarchical clustering as illustrated in Figure 3.

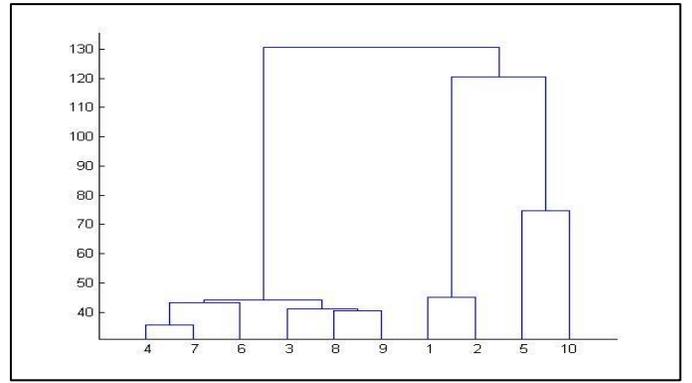

**Figure 3:** Hierarchical Clustering of Data Event Windows, vertical axis distance, horizontal axis cluster number.

Algorithm 1 is used to generate the clustering dendrogram as illustrated in Figure 3. Algorithm 1 implements a modified hierarchical agglomerative clustering algorithm that merges clusters until a minimum distance threshold between clusters is reached or all the clusters are merged to only one. When a minimum distance threshold is used, the algorithm ensures maximum partitioning of data into feature-rich clusters and increases the probability that the top-tier clusters contain a full attack signature feature set.

**Algorithm 1** takes as input (1) $SSig_{attack}$, the set of attack signatures, each of which consists of several features, (2) $\Phi$ with the $N$ samples, where each sample is a row vector in $\Phi$, and (3) $\delta$, the distance threshold to stop cluster merging. This algorithm groups the samples in one event window to $c$ clusters (Lines 3-8). Then, for each cluster, it finds the attack signature that has the highest probability to match the cluster's features (Lines 10-21). The signatures and matching probabilities for all the clusters are put into *SSig* and *PProb* respectively. This algorithm outputs a triple (*C*, *SSig*, *PProb*) where *C[i]* is the $i$-th derived cluster, *SSig[i]* contains the attack signature with the highest probability (in *PProb[i]*) to match *C[i]*'s features.

In this algorithm, *c* represents the total number of clusters found so far, and is initialized to 0 (Line 2). $D_i$ is a cluster that is being processed. Initially, $D_i$ is initialized to contain the *i*-th sample in one event window. *C* is the set of clusters finally derived, and is initialized as an empty set.

**Algorithm 1:** *SignatureMatchProb(SSig$_{attack}$, Φ, δ)*

```
1  begin
2  initialize c = 0; C={}; D = {D_i, …, D_N} where D_i is the i-th sample;
3  do /*merge cluasters*/
4      c = c + 1
5      find the two nearest clusters D_i and D_j from D
6      merge D_i and D_j to a new cluster and insert the new cluster to C
7  until dist(D_i, D_j) > δ
8  end do
9  i = 0; PProb={}; SSig={};
    /* for each cluster, find the attack signature
       which matches this cluster's features with the highest probability*/
10 for each cluster D_i (∈ C) do
11     k=0; maxProb_i=0; maxProbSig_i={};
12     for each attack signature S_k (∈ SSig_attack)
13         H_i = extract features from D_i that also exist in S_k
14         N_i = # of H_i feature with conditional entropy higher than H(F)
15         N_k = # features in S_k
16         if(N_i/ N_k> maxProb_i)
17             maxProb_i= N_i/ N_k
18             maxProbSig_i = S_k
19         end if
20         k = k + 1
21     end for
22     PProb = PProb ∪ {maxProb_i}
23     SSig = SSig ∪ {maxProbSig_i}
24     i = i+1;
26 end for
27 return (C, SSig, PProb);
28 end
```

Lines 3-8 merges samples to *c* significant clusters. This cluster merging process stops when the minimum distance between two nearest clusters exceeds the distance threshold *δ*. In finding the nearest clusters from all the existing ones, both Ward and complete linkage methods can be utilized. Past research [20] showed that the complete linkage method Eq. (10) yields the best ratio between the within group sum of squares (WGSS) and between groups sum of squares (BGSS); thus, indicating tighter grouping between inter-cluster members and optimal cluster-to-cluster spacing.

$$\text{dist}(D_i, D_j) = \max\{d(x_i, x_j): x_i \in D_i, x_j \in D_j\} \quad (10)$$

For each cluster $D_i$ out of the *c* clusters in *C*, Lines 10-21, calculate the probability that its features match every attack signature. The details of the signature matching are omitted in the algorithm for simplicity purpose. We discuss the details here. Identifying feature matches is performed by measuring the entropy for each feature within an event window. As Figure 4 indicates, the entropy of an individual significant feature $f_j$ increases when attack suite traffic is injected into the network.

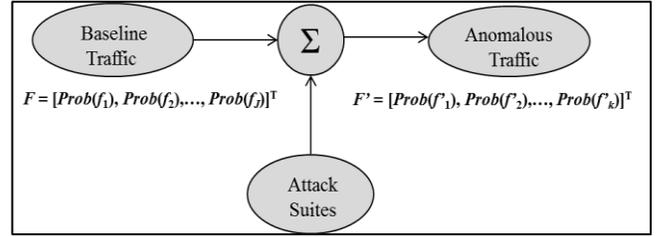

**Figure 4:** Features and entropy relationships.

The average entropy for features in the baseline traffic $H(F)$ is

$$H(F) = - \sum_{j=1}^{Nf} Prob(f_j) \log Prob(f_j) \quad (11)$$

The average conditional entropy for features in anomalous traffic is

$$H(F'|f'_k) = -\sum_{j=1}^{Nf} Prob(f_j|f'_k) \log Prob(f_j|f'_k) \quad (12)$$

It is important to only classify valid clusters. Only clusters with active feature frequencies greater than 2% of the total number of samples in an event window become candidates for classification. In addition, of the candidate clusters those with the largest cophenetic distances and highest inconsistency factor are selected for feature entropy comparisons.

A feature $f'_k$ appearing in anomalous traffic is significant if $H(F'/f'_k)$ is greater than $H(F)$. Significant features are subsequently used in determining the probability that a cluster is associated with a specific attack suite

|     | $D_1$ | $D_2$ | $D_5$ | $D_9$ |
|-----|-------|-------|-------|-------|
| $f_1$ | x | x | x | x |
| $f_2$ | x | x | x | x |
| $f_3$ |   | x | x | x |
| $f_9$ | x |   |   | x |

**Figure 5**: Attack Signature Matching Probabilities

Let us look at an example for signature matching probabilities. Assume that we are given an attack signature $S_k$ (∈ $SSig_{attack}$) which consists of four features, $f_1, f_2, f_3, f_9$, i.e., $S_k$ ={$f_1, f_2, f_3, f_9$}. Algorithm 1 finds clusters and extracts the features that exist in $S_k$. As shown in Figure 5, for candidate clusters $D_1, D_2, D_5$ and $D_9$, the algorithm founds matching features $H_1$ ={$f_1, f_2, f_9$}, $H_2$ ={$f_1, f_2, f_3$}, $H_5$ ={$f_1, f_2, f_3$}, and $H_9$ ={$f_1, f_2, f_3, f_9$}. They all have high (>=75%) feature matches to $S_k$. Among these four clusters, $D_9$ has the highest feature match probability (100%). I.e., all the four features in the attack signature $S_k$ exist in cluster.

After calculating the signature matching probability (Lines 13-16), the attack signature with the highest feature-matching probability is recorded by *maxProbSig$_i$* (Line 18) and the matching probability is recorded by *maxProb$_i$* (Line 17). When the highest matching probability *maxProb$_i$* and the matching attack signature *maxProbSig$_i$* for each cluster is found, they are put into sets *PProb* and *SSig*, respectively (Line 22-23).

*E. Calculate Assurance Level for a Path*

Once the set of clusters *C* are derived and probabilities of the presence of specific signatures in those clusters (*PProb* and *SSig*) are calculated, the path information assurance factor $I_i$ for network path $P_i$ is calculated using Eq. (13) and Eq. (14). Based on domain specific cyber security threat models [30], for each cyber threat level $W_i$, and a corresponding traffic threat signature $S_i$ present in an event window, the likelihood of cyber threat signatures being present is high if both $W_i$ and $Prob(S_i)$ are high. For a path $P_i$, which consists of *c* clusters (discovered in the previous step), we can sum up the threat for each cluster (Eq. (13)). Then the information assurance factor $I_i$ for $P_i$ is derived using Eq. (14):

$$O_f = \sum_{i=1}^{c} W_i \cdot Prob(S_i) \quad (13)$$

$$I_i = \frac{1}{O_f} \quad (14)$$

Algorithm 2 calculates the information assurance factor *I* for a path by utilizing Eq. (13) and (14).

---

**Algorithm 2:** *PathInfoAssuranceLevel(C, SSig, PProb, W$_r$)*

1 **initialize** *O* = 0; *i*=1;
2 **begin**
3   **for** the *i*-th cluster $D_i$ in *C*
4     Calculate its corresponding threat level $W_i$ using
        its attack signature $S_i$ ($\in$ *SSig*) and threat level $W_r$
5     Get its highest feature-matching probability *Prob$_i$* ($\in$ *PProb*)
6     *O* = *O* + $W_i$ × *Prob$_i$* /* According to Eq. (12) */
7     *i* = *i* + 1
8   **end for**
9   *I* = 1/*O* /* According to Eq. (14) */
10   **return** *I*
11 **end**

---

The path information assurance factor $I_i$ is calculated in Line 9.

*F. Summary of Methodology*

As mentioned in Section III.A, a total of 19 features were extracted from the packet records. Network traffic, which consists of the packet records, is partitioned into 30-minute event windows. For each 30-minute event window, a traffic feature frequency matrix *Φ* is extracted to contain the samples of the 19 features.

Algorithm 3 summarizes the complete algorithm to calculate the information assurance measurement *I* for each event window. It takes three parameters as input. The first parameter is the $N \times N^f$ sample matrix *Φ* for an event window. Its traffic data is composed of the baseline traffic and anomalies. The second parameter is the set of attack threat signature *SSig$_{attack}$*. It consists of *S* signatures. The third parameter is the threat level *W$_r$*.

---

**Algorithm 3:** *PathInfoAssurance(Φ, SSig$_{attack}$, W$_r$)*

1   $U_v \leftarrow$ *GetSensingMatrix(N, M)*
2   *Y* $\leftarrow$ *CSSample(U, M, Φ)* /* Section III.A*/
3   {$Z_1, ...., Z_{Nf}$} $\leftarrow$ *DetectAnomalies(Φ, Y)* /* Section III.B*/
4   **if** $\exists Z_i (\in \{Z_1, ...., Z_{Nf}\})$ *s.t.* $\|Z_i\|_2^2 > Q_\beta$ **then**
5     (*C*, *SSig*, *PProb*) $\leftarrow$ *SignatureMatchProb (Φ, SSig$_{attack}$, δ)* /*Alg. 1*/
6     *I* $\leftarrow$ *PathInfoAssuranceLevel(C, SSig, PProb, W$_r$)* /*Alg. 2 */
7     *Store(Y)*
8   **else**
9     *Store(Y)*
10   **end if**

---

The algorithm works as follows. In Line 1, the sampled data (feature frequency matrix) *Φ* is used to generate a CS sensing matrix $U_v$ (, which is a $M \times N$ matrix). In Line 2, *Φ* and the sensing matrix $U_v$ are multiplied to produce *Y*, an $M \times N^f$ matrix using Eq. (3). In Line 3, the volume-based anomaly detection is performed on each column of the *Y* matrix, and the corresponding prediction error $\|z\|_2^2$ is returned. In Line 4, if there exists any feature's prediction error $\|z\|_2^2 > Q_\beta$, it means that there is the likelihood that attack signatures are present in *Φ*. Then, *Φ* is analyzed for the presence of attack signatures (Line 5). Otherwise, the event window that produces *Φ* is determined to have a low probability of containing malicious content and is stored in compressed form for possible future analysis. In Line 5, the signature-matching component is called to determine the probability of the presence of specific attack signatures in this event window. It produces the attack signatures *SSig*, data clusters *C*, with highest signature presence of *PProb*. In Line 6, the information assurance value *I* for a path $P_i$ is calculated using the set of signatures *SSig*, and their corresponding matching probability (in *PProb*) to attack signatures.

*G. Complexity and Efficiency Gains*

The overall computational complexity of *PathInfoAssurance* expressed in Big *O* notation is as follows. Let *N* be the number of samples, *θ* be the number of non-sparse components, *c* be the number of clusters, *S* be the number of attack signatures, and $M = \mathcal{O}(\theta \log N)$.

| | |
|---|---|
| *GetSensingMatrix* | $\mathcal{O}(N^2 \log N)$ |
| *CSSample* | $\mathcal{O}(N^{2.373})$ |
| *DetectAnomalies* | $\mathcal{O}(M^3)$ |
| *SignatureMatchProb* | $\mathcal{O}(SN^2 \log N)$ |
| *PathInfoAssuranceLevel* | $\mathcal{O}(c)$ |

The following assumptions are considered when performing complexity analysis:

1. The number of signatures *S* can grow very large.
2. *DetectAnomalies* and its predecessors must always be run in order to detect zero-day attack behaviors within an event window.
3. The accuracy of *DetectAnomalies* is assumed high enough that *SignatureMatchProb* is executed only when anomalies are detected.

Taking into consideration that *M* and *c* are small while *N* is very large, the computational complexity lies primarily in *GetSensingMatrix*, *CSSample*, and *SignatureMatchProb*. As it is assumed that *DetectAnomalies* and its predecessors must process each event window, the principal savings come when no anomalies are detected and it is subsequently unnecessary to call *SignatureMatchProb*.

## IV. RESULTS

In this section, Section IV.A presents the strategies to collect traffic data and analyzes the characteristics of the network traffic. Section IV.B discusses the effect of applying the CS technique. Section IV.C then shows the accuracy of our presented approach. Section IV.D explains the gains in running time of our approach.

### A. Characterization of Sample Data

The goal of this research was to accurately model threats encountered by modern cloud service providers and clients. The most often used data sets, DARPA/Lincoln Labs packet traces [26], [27] and the KDD Cup data set derived from them [28], are found to be inadequate for such modeling as they are both over a decade old and do not contain modern threats. These data sets containing synthetic data that do not reflect contemporary attacks, and have been studied so extensively that members of the intrusion detection community find them to be insignificant [29]. For these reasons, the data sets used in this research consists of contemporary cloud service provider attacks generated on large scale test networks.

In order to establish the ground truth to evaluate the accuracy of anomaly detection, we conducted an analysis of the traffic in a baseline event window *B*, which is free of attacks, and the same window's traffic *Φ*, which is injected with attack data. In particular, the baseline traffic event windows *B* without anomalies were fully sampled and descriptive statistics (e.g., mean, standard deviation, correlations) were calculated. Then, router and host node attacks were singly launched on the network where they were fully sampled. Descriptive statistics and signatures for each attack were calculated. This information established the ground truth for later analysis. The router and host attacks were injected into the baseline data in a random Poisson distribution to form anomalous event windows *Φ*. Compressed event windows were assembled via compressed sensing of individual anomalous event windows.

We analyzed the characteristics of the normal baseline traffic data *B* and the traffic data with injected malicious traffic *Φ*. This analysis was conducted prior to CS sampling and subsequent path information assurance level determination. Examples of 30-minute event windows for *B* and *Φ* are shown in Figures 6 and 7. For the event window *B* with normal traffic (i.e., baseline), we plotted in Figure 6 the percentage of the frequencies of principal features.

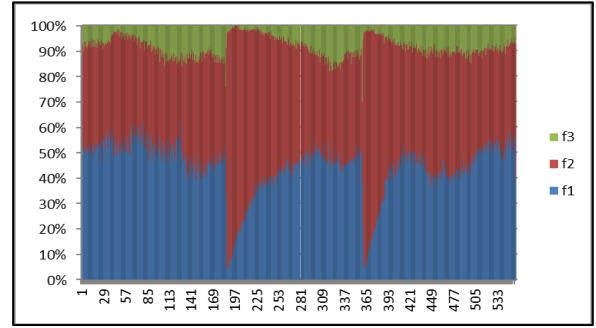

**Figure 6:** Baseline Data Free of Attacks (*B*), vertical axis percentage, horizontal samples.

These features are ICMP redirect (feature $f_1$), HTTP reset (feature $f_2$), and synchronization (feature $f_3$) packets.

**Table 2** Principal Features for Baseline Network Traffic

| Feature | Indicator |
|---|---|
| $f_1$ | ICMP Redirect |
| $f_2$ | TCP http [RST] |
| $f_3$ | TCP http [SYN, ACK] |

The principal features for the baseline data (in normal traffic) all showed a large measure of variance over the sampling period. The large variance measurements indicate that these features can be adequately sensed during the CS sampling process. For the traffic data *Φ* in an event window with anomalous behavior, we plotted the percentage of major features in Figure 7.

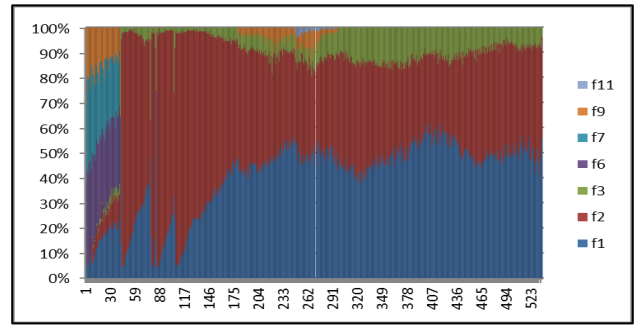

**Figure 7:** Baseline Data Plus Injected Attacks (*Φ*), vertical axis percentage, horizontal samples.

Figure 7 shows that the number of significant features in $\Phi$ increased, due to the addition of abnormally high Border Gateway Protocol (BGP), HTTPS, OSPF, and SSH protocol traffic resulting from router and host attack network traffic. Figure 7 also shows that the variance for each feature is also relatively high which indicate that CS sampling can effectively capture data patterns.

**Table 3** Principle Features for Network Traffic with Attacks

| Feature | Indicator |
|---|---|
| $f_1$ | ICMP Redirect |
| $f_2$ | TCP http [RST] |
| $f_3$ | TCP http [SYN, ACK] |
| $f_6$ | TCP bgp [RST, ACK] |
| $f_7$ | TCP ospf-lite [RST, ACK] |
| $f_9$ | TCP https [RST, ACK] |
| $f_{11}$ | TCP ssh [RST, ACK] |
| $f_{12}$ | TCP telnet [RST, ACK] |

Attack signature data was characterized prior to processing. A router attack signature is shown in Figure 8, which indicates that features $f_2, f_6, f_7,$ and $f_9$ are significant features.

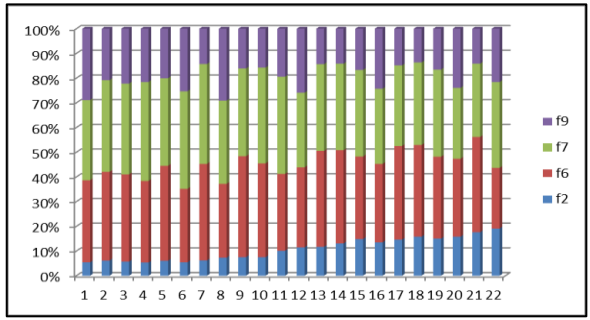

**Figure 8:** Router Attack Signature,
vertical axis percentage, horizontal samples.

### B. Effect of Compressed Sensing (CS)

Using FFT generated frequency components of $\Phi$ to generate the sampling matrix $U_v$, the restricted isometric constant (RIC) $\delta_k$ with $\theta$ representing non-sparse components, was kept very small. Keeping $\delta_k$ small guaranteed good linear transformation of $\Phi$ and high orthogonality of columns in $U_v$.

$$(1 - \delta_k)\|\Phi\|_2^2 \leq \|U_v\Phi\|_2^2 \leq (1 - \delta_k)\|\Phi\|_2^2 \quad (15)$$

The dimension $M$ was optimally determined by using Eq. (6), and the error rate was measured. Because $U_v$ is highly orthogonal, the geometry preserving properties allow for the detection of volumetric anomalies in the residual subspace. The largest eigenvalues constituting 90% of the spectral power. With a probability of 1- *conf* where *conf* ($\in [0,1]$) is the confidence interval, the changes in the eigenvalues between residual ($\mathbf{3}_i$) and principle ($\lambda_i$) subspaces equal to $|\lambda_i - \mathbf{3}_i|$ and are bound by:

$$|\lambda_i - \mathbf{3}_i| \leq 4\sqrt{2\lambda_1}\left(\sqrt{\frac{n_v}{M}} + \sqrt{\frac{2\ln\frac{1}{conf}}{M}}\right) \quad (16)$$

Where $\lambda_1$ is the largest eigenvalue found in the residual subspace, $i = 1,.., n_v$.

The false alarm rate $\Delta F$ is bound by:

$$\Delta F \leq O\left(\sqrt{\frac{n_v}{M}} + \sqrt{\frac{2\ln\frac{1}{conf}}{M}}\right) \quad (17)$$

Traditionally, the confidence threshold 90% is used which makes the $\sqrt{2\ln\frac{1}{conf}}$ term small when compared to $\sqrt{\frac{n_v}{M}}$. Thus, a smaller $M$ increases the probability of a false alarm and also increases the compression error rate [10]. The obvious advantage in using a smaller $M$ was a lower computational overhead. Accordingly, $M$ was chosen such that the intrinsic sparseness of $\Phi$ represented by $\varepsilon * \sqrt{N}$ in Eq. (6) yields the lowest compression error with $\varepsilon * \sqrt{N} \ll M \ll N$.

The optimal derivation of the constant ε in Eq. (6) was achieved by identifying those feature components of *x* that had the highest variance and magnitudes. The value of ε was found to be directly proportional to the number of active features. The compression mean squared error (MSE) was verified by measuring the error contained in convex optimized reconstruction.

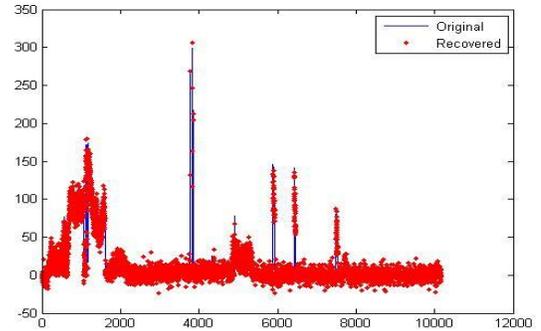

**Figure 9:** CS Reconstruction, Baseline + Injected Attacks ($G \leftarrow Y$), vertical axis frequency, horizontal samples.

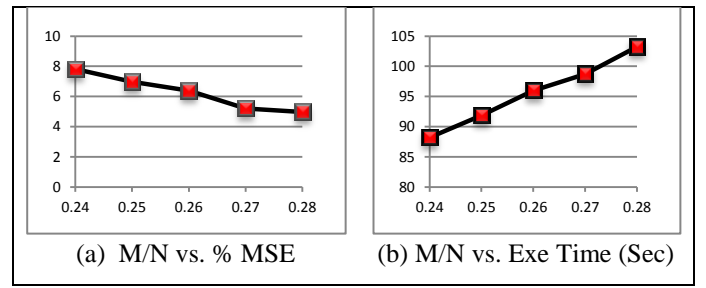

(a) M/N vs. % MSE      (b) M/N vs. Exe Time (Sec)

**Figure 10**: Compression ratio *M/N* versus MSE where *N* is the number of samples (without CS) and *M* is the number of sam-

ples (with CS) (a) Recovered data fidelity and (b) Execution Time.

Figure 11 shows the $\|z\|_2^2$ values calculated from the baseline and anomalous event windows.

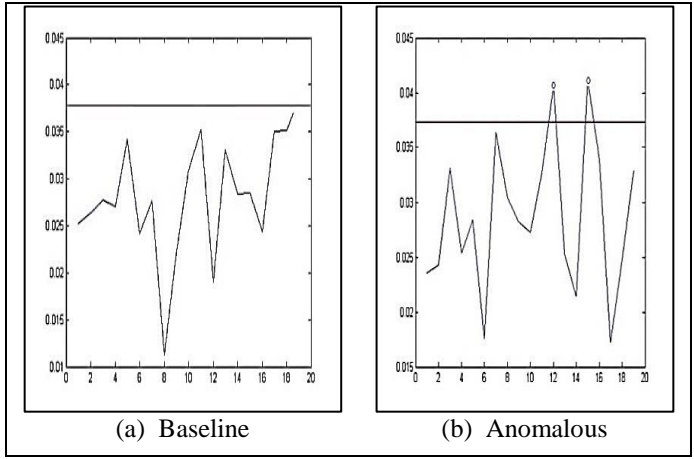

(a) Baseline　　(b) Anomalous

**Figure 11:** Detected Anomalies: $Q_B$ – horizontal line Q-Statistics threshold; horizontal axis represents network traffic features; vertical axis $\|z\|_2^2$.

The $\|z\|_2^2$ values for the baseline network traffic features are all less than the $Q_B$, Q-Statistics threshold indicating an absence of network attack traffic (a). Figure 11(b) shows $\|z\|_2^2$ values which is greater than $Q_B$ indicating an increase in the number and magnitude of network traffic attack features and a high possibility that network attack traffic is present in the event window.

Table 4 shows the relationships exhibited between optimally derived values for *M*, associated error, false positive alarm rate, and execution time for a 24 hour testing period with attack suites 1-3 randomly injected in a Poisson distributed. Each of the 3 attack suites discussed in Table 1 was analyzed resulting in an average effective anomaly detection performance greater than 93%.

**Table 4** Anomaly Detection Accuracy

| Attack Suite | No. of instances | Detected | False Pos. | False Neg. |
|---|---|---|---|---|
| 1 | 5269 | 4894 | 12 | 10 |
| 2 | 8020 | 7575 | 24 | 27 |
| 3 | 2920 | 2802 | 12 | 5 |

*C. Overall Accuracy*

The following table summarizes the observed accuracy of Algorithms 1, 2, and 3 in correctly detecting event window anomalies and in performing subsequent classifications.

**Table 5** Cluster Signature Probability Accuracy

| Attack | No. of | No. of | Classified | Avg. Threat |
|---|---|---|---|---|
| Suite | Attack instances | Forwarded Attack instances | Instances with > 90% Confidence | Measure % Accuracy |
| 1 | 5269 | 4894 | 4635 | 87.96% |
| 2 | 8020 | 7575 | 7194 | 89.70% |
| 3 | 2920 | 2802 | 2703 | 93.14% |

Out of a total of 16,209 attack instances, residual subspace anomaly detection correctly sensed 94.21% and forwarded those attack instances for subsequent classification. From the set of forwarded attack instances, Algorithm 1 correctly classified 95.16% with an average confidence greater than 90%. In total, the *DetectAnomalies-SignatureMatchProb* chain identified an average of 90.27% of all attack instances injected along network paths with an average probability of correct match greater than 90%.

*D. Efficiency*

Figure 12 illustrates the efficiency gains derived when *SignatureMatchProb* is not called. The vertical axis represents the execution time in seconds while the horizontal axis represents the ID of the data sets presented to the system. To illustrate the efficiency gains, each data set was presented to the chain initially without removing non-anomalous event windows. Then the same data set was presented to the chain, but allowing non-anomalous windows to be dropped prior to classification. In the case of first data set, two out of the four event windows were dropped, which leads to the corresponding efficiency gain of 50% in the signature classification phase. Similar efficiency gains were recorded for all data sets.

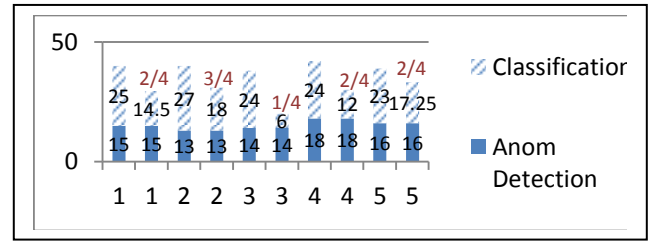

**Figure 12:** CS Anomaly detection and Signature Classification: horizontal axis represents tested data sets; vertical axis is run time in seconds.

Summarizing, an acceptably high percentage of attack instances were detected by anomaly detection (94.21%) of which 95.16% of these forwarded instances were associated with attack suite signatures with high confidence. Overall efficiency gains of over 50% were observed when non-anomalous event windows are dropped prior to classification. Additionally, using this unique combination of methods, the information assurance factor $I_i$ on any path $P_i$ was derived with greater than 90% confidence.

## V. CONCLUSION

In this paper, we studied the problem of determining the information assurance level for different paths in multipath networks. We showed it was possible to intelligently sense and quantify threats along individual paths with a high degree of confidence. In the process, we devised a novel approach that combines optimal network data sampling (CS) with residual subspace PCA anomaly detection and probabilistic signature-based intrusion detection. This methodology was shown to simultaneously compress sampled network traffic and reduce data dimensionality by filtering out non-contributing network traffic features. On the compressed and dimension-reduced data set, our approach efficiently performs path threat detection and classification. This approach increases the efficiency and data handling capabilities of both the anomaly and signature-based detection algorithms.

We also derived a theoretical multipath *SQoS* relation, Eq. (1). This relationship allows for the dynamic adjustment of security measures along each path and maintains the overall throughput at the same time. The determination of the security measures using our newly developed approach solves the most technically challenging portion of the multipath *SQoS* relation Eq. (1). Our approach and the multipath SQoS relations lay a solid foundation for the future expansion of adaptive multipath security approaches.

## AUTHORS

**James Obert** the principle author, joined Sandia National Labs in 2009 as a computer scientist and is actively involved in dynamic network defense and trusted computing research and development. Before joining Sandia, James engaged in cyber security R&D as a computer scientist at Booz Allen Hamilton, HP, IBM, and NASA. James is currently completing his PhD at New Mexico State University, and received a B.S.E.E. at the University of Texas, M.S.E.E. at New Mexico State University, and M.S.C.S at California State University.

**Huiping Cao** received her Ph.D. in Computer Science from the University of Hong Kong. She is an Assistant Professor in Computer Science at New Mexico State University (NMSU). Dr. Cao's research interests are in the areas of data mining and databases. She has published data management and data mining articles in highly competitive venues. Dr. Cao has served at the editorial board for Journal on Data Semantics (JoDS), as reviewer for peer-reviewed journals, and as program committee member for many international conferences and workshops.